\documentclass[fleqn,useAMS,usenatbib]{mnras}

\usepackage[T1]{fontenc}
\usepackage{ae,aecompl}

\usepackage{pdflscape}
\usepackage{graphicx}	
\usepackage{amsmath}	
\usepackage{amssymb}	
\usepackage{color}

\newcommand{\kms}{\hbox{${\rm km\;s}^{-1}$}}
\newcommand{\hi}{H~\textsc{i}}   

\newcommand{\Msun}{\ensuremath{M_{\sun}}}
\newcommand{\Mstar}{\ensuremath{M_{\star}}}
\newcommand{\logmstar}{\ensuremath{\log \, (M_{\star}/M_{\sun})}}
\newcommand{\logmstarshort}{\ensuremath{\log \, M_{\star}}}
\newcommand{\logre}{\ensuremath{\log \, \re}}

\newcommand{\MHI}{\ensuremath{M_{\mathrm{H} \textsc{i}}}}

\newcommand{\fgas}{\ensuremath{f_{\mathrm{gas}}}}
\newcommand{\logfgas}{\ensuremath{\log f_{\mathrm{gas}}}}

\newcommand{\sfourg}{S\ensuremath{^{4}}G}
\newcommand{\avis}{\ensuremath{a_{\mathrm{vis}}}}
\newcommand{\logavis}{\ensuremath{\log a_{\mathrm{vis}}}}

\newcommand{\re}{\ensuremath{R_{e}}}
\newcommand{\msepred}{\ensuremath{\mathrm{MSE}_{\mathrm{pred}}}}

\defcitealias{erwin18}{Paper~I}
\defcitealias{scipy}{Jones et al.\ 2001}

\title[Bar Sizes in Spirals]{What Determines the Sizes of Bars in Spiral Galaxies?}
\author[P. Erwin]{Peter Erwin$^{1,2}$\thanks{E-mail: erwin@mpe.mpg.de} \\
$^{1}$Max-Planck-Insitut f\"{u}r extraterrestrische Physik, Giessenbachstrasse, 85748 Garching, Germany \\
$^{2}$Universit\"{a}ts-Sternwarte M\"{u}nchen, Scheinerstrasse 1, D-81679 M\"{u}nchen, Germany}

\date{Accepted XXX. Received YYY; in original form ZZZ}

\pubyear{2019}

\begin{document}
\label{firstpage}
\pagerange{\pageref{firstpage}--\pageref{lastpage}}
\maketitle

\begin{abstract} 

I use volume- and mass-limited subsamples and recently published data
from the Spitzer Survey of Stellar Structure in Galaxies (\sfourg) to
investigate how the size of bars depends on galaxy properties. The known
correlation between bar semi-major-axis $a$ and galaxy stellar mass (or
luminosity) is actually \textit{bimodal}: for $\logmstar \la 10.1$, bar
size is almost independent of stellar mass ($a \propto \Mstar^{0.1}$),
while it is a strong function for higher masses ($a \propto
\Mstar^{0.6}$). Bar size is a slightly stronger function of galaxy
half-light radius \re{} and (especially) exponential disc scale length
$h$ ($a \propto h^{0.8}$). Correlations between stellar mass and galaxy
size can explain the bar-size--\Mstar{} correlation -- but only for
galaxies with $\logmstar \la 10.1$; at higher masses, there
is an extra dependence of bar size on \Mstar{} itself.
Despite theoretical arguments that the presence of gas can affect bar
growth, there is no evidence for any residual dependence of bar size on
(present-day) gas mass fraction. The traditional dependence of bar size
on Hubble type (longer bars in early-type discs) can be explained as a
side-effect of stellar-mass--Hubble-type correlations. Finally, I show
that galaxy size (\re{} or $h$) can be modeled as a function of stellar
mass and both bar presence and bar size: barred galaxies tend to be more
extended than unbarred galaxies of the same mass, with larger bars
correlated with larger sizes.

\end{abstract}


\section{Introduction} 

The majority of massive disc galaxies in the local universe possess
stellar bars \citep[e.g.,][and references therein]{erwin18}, which are
important for understanding the evolution of disc galaxies for a number
of reasons. The mere presence of a bar can be taken as an indication of
certain key developments in galaxy evolution, such as the appearance of
dynamically cool discs \citep[e.g.,][]{sheth12}. Bars themselves have
also been implicated as possible -- even essential -- mechanisms for the
quenching of star formation in discs
\citep[e.g.,][]{gavazzi15,spinoso17,khoperskov18}, the growth of
pseudobulges, transformations of dark matter halos
\citep[e.g.,][]{weinberg02,holley-bockelmann05,weinberg07b,dubinski09},
and the source of variations in the $M_{\rm BH}$--$\sigma$ correlations
of supermassive black holes \citep[e.g.,][]{brown13}.

The simplest, least ambiguous, and most easily measured characteristic
of bars is their size. (Other important characteristics, such as
``strength'' and rotational pattern speed, are harder to define
unambiguously or much more difficult to measure.) Models of bar
evolution can make at least semi-quantitative predictions about how much
bars should \textit{grow} after they form, depending on factors such as
the gas content of discs and the degree of interaction with dark-matter
halos. For example, a number of studies have shown that bars can slow
down and grow in length, especially if they exchange angular momentum
with bulges and dark-matter halos
\citep[e.g.,][]{hernquist92,debattista98,debattista00,athanassoula02b,
athanassoula03,algorry17}. The latter effect can in turn significantly
reshape the halos.

Conversely, several models have found that even moderate amounts of gas
in discs (e.g., gas mass $\ga 10$\% of the disc mass) can lead to
delayed bar formation and slower growth \citep[e.g.,][]{berentzen98,
berentzen07,bournaud05c,villa-vargas10,athanassoula13}, in part because
bars can \textit{gain} angular momentum from gas located within
corotation, with the gas driven in towards the center of the galaxy,
possibly fuelling bulge or pseudobulge growth and nuclear activity.
These studies suggest that, all other things being equal, we might
expect shorter bars in more gas-rich galaxies, either because the bars
formed more recently and have had less time to grow, or because their
growth has been reduced or suppressed. (Of course, there is the
complicating factor of how much \textit{present-day} gas content
might or might not reflect the gas content several Gyr ago, when
the bar was forming or (trying) to grow.)

Finally, the sizes of bars turn out to be important for interpreting
studies of bar \textit{frequency} in large samples, including those at
intermediate and high redshifts. This is for the simple reason that
physically smaller bars are harder to detect at larger distances. If
there are differences in bar sizes as a function of some host-galaxy
parameter, then differential visibility may produce a false signal of
differential bar presence. In \citet[][hereafter Paper~I]{erwin18}, I
argued that just such an effect was responsible for claims of low bar
frequencies -- and a bar frequency that monotonically decreased to lower
masses -- in most large, SDSS-based samples. Specifically, if bar size
correlates with stellar mass (as indeed it does;
Section~\ref{sec:barsize-mass}), then in resolution-limited samples,
bars will be more easily detected in higher-mass galaxies, and will thus
appear to be less common in lower-mass galaxies than is actually the
case. I also pointed out that this could affect attempts to measure bar
frequency as a function of redshift, since constant-angular-resolution
images (e.g., from \textit{Hubble Space Telescope} observations in a
given filter) potentially make it harder to detect bars at higher
redshifts.

The preceding argument about bar size and bar detectability in imaging
studies was made in part by looking briefly at the trend of bar size
with stellar mass in the \sfourg{} sample (e.g., Section~4.2 of
\citetalias{erwin18}), but also by simply using the \sfourg{} sample as
an empirical ``parent distribution'' for simulating SDSS-based (and
high-redshift) surveys. In this paper, I return to the question of how
bar size relates to galaxy parameters in a more general sense, by
looking more carefully at how bar size in the \sfourg{} sample does (or
does not) depend on stellar mass, galaxy size (half-light radius \re),
disc size (exponential scale length $h$), gas mass fraction, and Hubble
type.

After reviewing the definition of the \sfourg-based sample and the
additional data sources I use in Section 2, I discuss the strong -- and
curiously \textit{bimodal} -- dependence of bar size on stellar mass in
Section 3.1, followed by discussions of the (even stronger) dependence
on galaxy size and (strongest of all) on disc size in Section 3.2. Since
galaxy and disc size are correlated with stellar mass, Section 3.3 looks
at the question of whether any of these individual bar-size relations
are merely side effects of the stellar-mass--galaxy/disc-size
correlation. The answer, interestingly, is that \textit{both} galaxy
size \textit{and} stellar mass help determine bar size: more massive
galaxies have larger bars than less massive galaxies of the same size.
Section~\ref{sec:other-params} turns to the issue of whether bar size is
related to (present-day) gas content or Hubble type, finding no evidence
for any correlation of bar size with either of those parameters -- once
their correlations with stellar mass are corrected for. Finally
Section~\ref{sec:discuss} discusses some implications for models of bar
and disc formation, and Section~\ref{sec:summary} summarizes things.

To aid in reproducibility, there is a Github repository containing data
files, code, and Jupyter notebooks for creating the figures and fits
from this paper; the repository is available at
\url{https://github.com/perwin/s4g_barsizes} and also at
\url{https://doi.org/10.5281/zenodo.86151029}.

\section{Sample and Data Sources}\label{sec:sample} 

The best local sample for assessing general bar properties is the
Spitzer Survey of Stellar Structure in Galaxies
\citep[\sfourg;][]{sheth10}, both because of its size and because of its
use of near-infrared imaging, which minimizes the possibility of missing
or mismeasuring bars due to the confusion introduced by dust extinction
and star formation. \citet{dg16a} defined a subsample of non-edge-on
disc galaxies, which takes the version of \sfourg{} with morphological
classifications by \citet{buta15} and then removes both elliptical
galaxies and disc galaxies with inclinations $> 65\degr$ -- the second
step being crucial for maximizing the detectability of bars -- leaving a
total of 1344 galaxies. Ten galaxies are missing distances and stellar
masses in \citet{munoz-mateos15}, and an additional twelve have very
uncertain distances (redshifts of $< 500$ \kms{} with no alternate
distance measurements) and/or optical diameters smaller than the
\sfourg{} limit. Removing these leaves a total of 1322 galaxies in what
I call the Parent Disc Sample.

As discussed in \citetalias{erwin18}, \sfourg{} suffers from
incompleteness in three ways. First, since the sample is
magnitude-limited, lower-luminosity (and thus lower-mass) galaxies drop
out as the distance increases (see, e.g., Figure~1 of Paper~I). I deal
with this by adopting a distance- and mass-limited subsample, using
galaxies with $\logmstar = 9$--11 and distances $\leq 30$ Mpc. 
\sfourg{} is also incomplete in terms of Hubble types -- specifically,
it is incomplete when it comes to S0 galaxies (and ellipticals), since
\hi{} radial velocities were used for determining whether or not
galaxies met the redshift limit for the sample. (Additional
\textit{Spitzer} observations of gas-poor elliptical and S0 galaxies
have been made -- e.g., \citealt{knapen14} -- but the relevant analysis
of these galaxies is not yet publicly available.) The simplest way to
deal with this is to exclude the S0 galaxies and concentrate on spirals
(and the small number of irregulars in the sample). The end result is
the Parent Spiral Sample (623 galaxies), with a total of 387 barred
galaxies. (The Parent Spiral Sample is equivalent to ``Sample 2m'' in
\citetalias{erwin18}, except that the latter included galaxies with
$\logmstar > 11$.) The identification of galaxies as barred or
unbarred is based on the classifications and measurements of
\citet{buta15} and \citet{herrera-endoqui15}; see
Section~\ref{sec:size-measurements}.

As explained below, I use galaxy \textit{size} measurements --
half-light radius \re{} and exponential disc scale length $h$ -- from
the 2D decompositions of \citet{salo15}. These do not exist for all the
galaxies in the Parent Spiral Sample, so for most of the comparisons, I
use a subset containing those galaxies which do have valid $\re$ and $h$
measurements; this is the Main Spiral Sample (588 galaxies), with 367
barred galaxies making up the Main Barred Spiral Sample. A summary
of the various samples and subsamples is given in Table~\ref{tab:samples}.

\begin{table}
\caption{Galaxy Samples}
\label{tab:samples}
\begin{tabular}{lrrl}
\hline
Name      & $N_{\rm tot}$  & $N_{\rm bar}$ & Notes\\
\hline
Parent Disc & 1322 & 744 &  \\
Parent Spiral & 623 & 387 &  $D < 30$, $\logmstar = 9$--11 \\
Main Spiral   &  588 & 367 & valid \re{} and $h$ only \\
\end{tabular}

\medskip

The main samples used in this paper. The Parent Disc Sample is based on
the low- and moderate-inclination disc-galaxy subsample of \sfourg{}
defined by \citet{dg16a}. The Parent Spiral Sample is a subset
with distances $< 30$ Mpc and $\logmstar = 9$--11. The Main Spiral
Sample excludes galaxies which do not have valid \re{} and $h$
measurements; see Section~\ref{sec:sample} for more details.

\end{table}

\subsection{Data Sources} 

Details for most of the data sources are given in \citetalias{erwin18};
I summarize things briefly here. Distances and stellar masses are taken
from \citet{munoz-mateos15},\footnote{These distances are based on
redshift-independent NED distances for 79\% of the total sample, and on
Hubble-flow distances for the rest, assuming $H_{0} = 71$ \kms{}
Mpc$^{-1}$.} overall galaxy orientations (disc position angles and
inclinations) are from \citet{salo15}, and bar-size and position-angle
measurements are from \citet{herrera-endoqui15}. Atomic (hydrogen) gas
masses are based on systematized Hyperleda \hi{} measurements; both the
Parent and Main Barred Spiral Samples are about 99\% complete in terms
of \hi{} measurements.\footnote{The gas masses do not include any
corrections for helium or metals.}

\subsubsection{Bar and Galaxy Sizes}\label{sec:size-measurements} 

Bar sizes for \sfourg{} barred galaxies \citep[as identified
by][]{buta15} were measured by \citet{herrera-endoqui15}. These include
multiple different measures of bar size; I use their ``visual''
semi-major axes $\avis$, since these are available for \textit{all} the
barred galaxies, while other measurements are missing for some galaxies
based on factors such as low S/N or strong star-formation in the images.
(Note that \citealt{dg16a} find that the \avis{} and
maximum-ellipticity-based sizes are consistent for galaxies that have
both.) Since deprojected sizes are given in \citealt{herrera-endoqui15}
only for galaxies with ellipse-fit measurements, I derive a consistent
set of intrinsic bar sizes by deprojecting the \avis{} measurements
using their bar position angle values and the disc position angles and
inclinations from \citet{munoz-mateos15}.

To help in the comparison of different fits, I assume a constant
fractional uncertainty of 10\% for bar sizes (0.044 in logarithmic
terms). As a partial justification for this, I note that \citet{hoyle11}
found that different GZ2 participants reproduced each other's (visual)
bar measurements to $\sim 10$\%.

Finally, I use two galaxy size estimates, both from the 2D fits to the
\sfourg{} 3.6\micron{} images by \citet{salo15}. The first is a global
half-light radius \re{} that comes from single-S\'ersic fits. The second
is an exponential-disc scale length $h$ from their multi-component fits.
When multiple exponential components were used in the fit, I take the
largest scale length. (As noted above, there are a total of 588 galaxies
with both \re{} and $h$ measurements, making up the Main Spiral Sample.)

\section{The Sizes of Local Bars in \sfourg{}}\label{sec:barsizes} 

\subsection{Bar Size as a Function of Stellar Mass}\label{sec:barsize-mass} 

\begin{figure}
\begin{center}
\hspace*{-3mm}\includegraphics[scale=0.46]{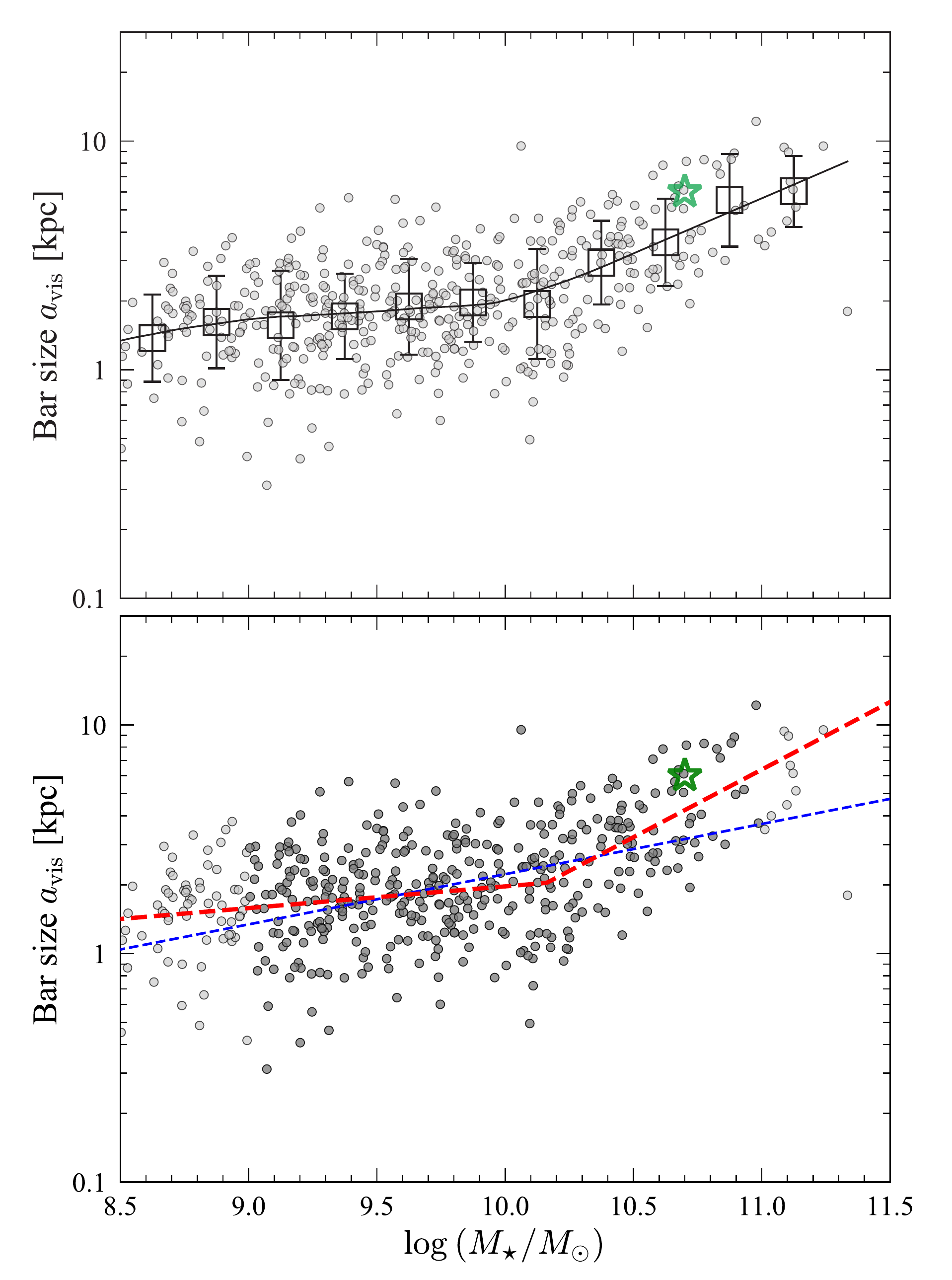}
\end{center}

\caption{Bar semi-major axis versus stellar mass for \sfourg{} spiral
galaxies with $D \leq 30$ Mpc. Top: data with LOESS fit (black line);
boxes indicate binned mean values, with error bars showing standard
deviations. Bottom: data with linear fit (thin, short-dashed blue line)
and broken linear fit (thick, long-dashed red line); in both cases, the
fit is to the Parent Spiral Sample ($\logmstar = 9$--11), indicated by
the darker data points. The green star in both panels indicates the
location of the Galaxy's bar in the plot, based on the estimate of
\citet{wegg15} and assuming $\Mstar = 5 \times 10^{10} \Msun$ (this
point is not used in any of the fits).}\label{fig:mstar-fit}

\end{figure}

\begin{figure}
\begin{center}
\hspace*{-2.5mm}\includegraphics[scale=0.46]{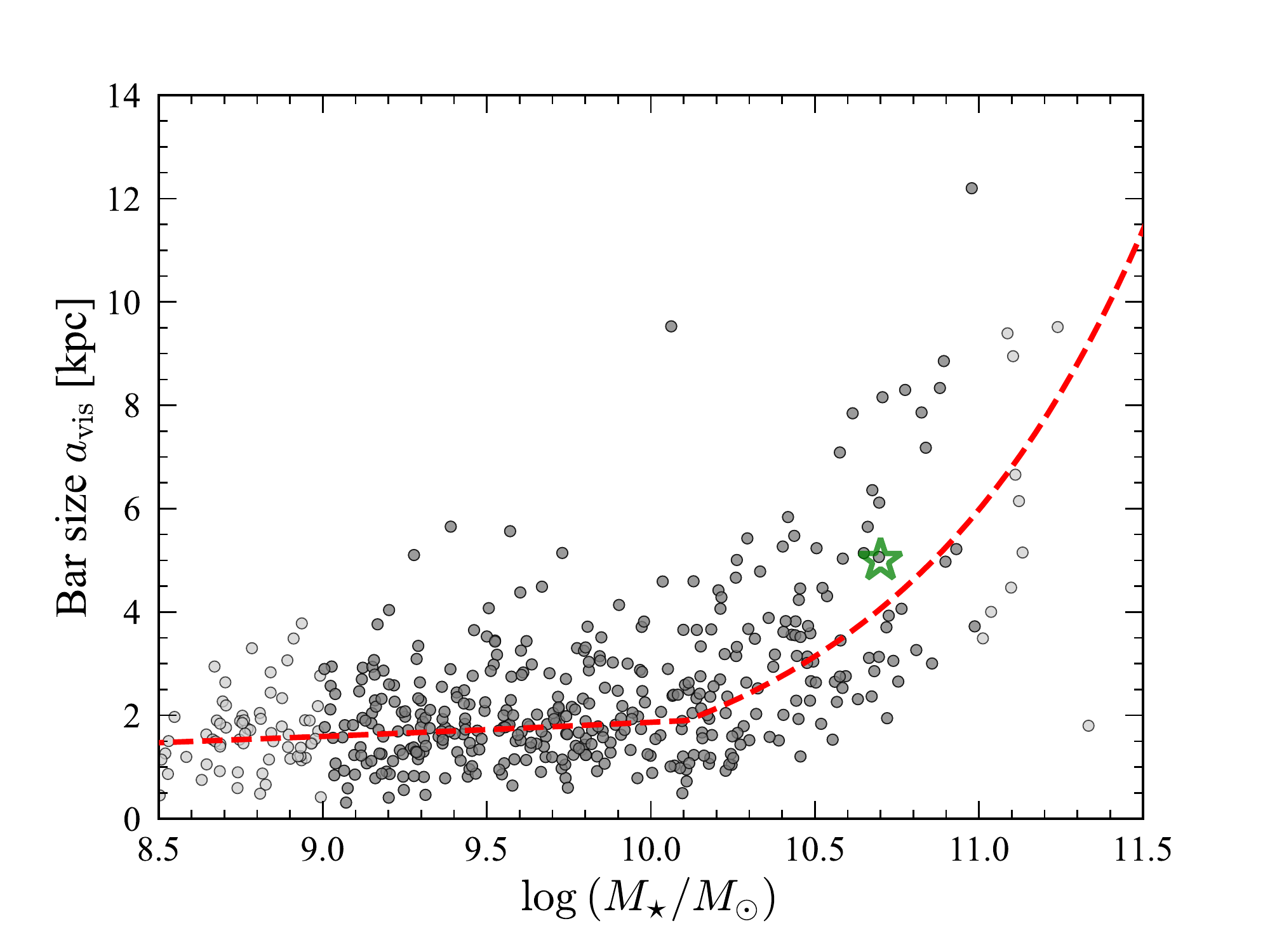}
\end{center}

\caption{As for the bottom panel in Figure~\ref{fig:mstar-fit}, but now showing bar sizes on
a linear scale.}
\label{fig:mstar-fit-linear}

\end{figure}

The oldest association between bar size and galaxy properties is the
observation that bar size scales with galaxy luminosity, as originally
pointed out by \citet{kormendy79a}. It makes sense to assume that such a
correlation reflects a more fundamental correlation with galaxy
\textit{stellar mass}, and \citet{dg16a} demonstrated this was indeed
the case for S4G galaxies (see their Fig.~20 and Table~5).

Figure~\ref{fig:mstar-fit} shows how (deprojected) bar size behaves as a
function of stellar mass for \sfourg{} galaxies in the Parent Barred
Spiral Sample. Close inspection suggests the trend actually has a
\textit{bimodal} quality: bar size is almost constant (but increases
slightly with stellar mass) for stellar masses $\la 10^{10} \Msun$, and
then increases more rapidly for higher masses.\footnote{This can also be
seen in the bottom panel of Fig.~20 of \citet{dg16a}.} In the upper
panel of Figure~\ref{fig:mstar-fit}, I include the result of a LOESS
(locally weighted regression) fit to the data \citep[using the code
of][]{cappellari13b-atlas3d-20,cappellari14} as a thin black line. (This
fit is inevitably biased near the ends of the distribution, because
there are fewer galaxies with masses below or above the limits.) This
reinforces the suggestion that the trend is bimodal: almost flat for
$\logmstar \la 10$, then changing to a steeper linear relation for
higher masses. (I also include in this figure a point for the Galaxy's
bar, which can be seen to have a length quite typical for a galaxy of
its stellar mass.) There is no systemic variation in the
\textit{spread} of bar size as a function of stellar mass; the error
bars in the upper panel of Figure~\ref{fig:mstar-fit} indicate that the
standard deviation of binned bar size does not change systematically
from low to high mass.

A simple way to quantify this bimodality is to fit the
bar-size--stellar-mass relation with a broken power law -- or,
equivalently, a broken-linear relation in the $\log \avis$--$\logmstar$
plane. This can be expressed as the following (with \avis{} in kpc and
\Mstar{} in solar masses):
\begin{equation}\label{eq:barsize1}
  \begin{aligned}
\log \avis & = & \alpha_{1} + \beta_{1} \logmstarshort \;\; \mathrm{if} \; \logmstarshort < M_{\rm brk}, \\ 
& = & \alpha_{2} + \beta_{2} \logmstarshort \;\; \mathrm{if} \; \logmstarshort > M_{\rm brk}
  \end{aligned}
\end{equation}
The constraint that the two linear pieces should match at $\Mstar =
M_{\rm brk}$ means that one of the parameters can be expressed in terms
of the others: e.g., $\alpha_{2} = \alpha_{1} + (\beta_{1} - \beta_{2})
M_{\rm brk}$. I fit this function to the data using a simple
least-squares minimization technique (the \texttt{curve\_fit} function
from the Python \texttt{scipy.optimize} package; \citetalias{scipy}).

The results of fits to the Parent and Main Spiral Samples are listed in
Table~\ref{tab:bar-size-fits}; the uncertainties on the best-fit
parameter values come from 2000 rounds of bootstrap resampling. Both
fits agree that the ``break mass'' is $\logmstar \sim 10.1$--10.2, with
a slope of $\sim 0.1$ for lower masses and $\sim 0.6$ for higher masses.
The dashed red line in the lower panel of Figure~\ref{fig:mstar-fit}
shows the fit using data from the Main Spiral Sample.

For an order of magnitude increase in galaxy mass, from $10^{9}$ to
$10^{10} \Msun$, the typical bar increases only $\sim 20$\% in
semi-major axis, from $\sim 1.5$ to $\sim 1.8$ kpc. But for larger
masses, the bar size increases dramatically ($\avis \sim \Mstar^{0.6}$):
the typical bar in a $10^{11} \Msun$ galaxy has a semi-major axis of
$\sim 5.9$ kpc. Figure~\ref{fig:mstar-fit-linear} illustrates this
increase by showing \textit{linear} bar size as a function of stellar
mass.

\begin{table*}
\begin{minipage}{126mm}
\caption{Fits to Bar Size versus Stellar Mass}
\label{tab:bar-size-fits}
\renewcommand{\arraystretch}{1.5}
\begin{tabular}{lcccc}
\hline
Sample & $\alpha_{1}$ & $\beta_{1}$ & $\beta_{2}$ & $\log \: (M_{\rm brk} / \Msun)$ \\
  (1)  & (2)          & (3)         & (4)         &        (5)  \\
\hline
Parent Spiral   & $-0.66^{+0.37}_{-0.41}$   & $0.10^{+0.04}_{-0.04}$    & $0.59^{+0.08}_{-0.09}$    & $10.16^{+0.07}_{-0.08}$   \\
Main Spiral     & $-0.42^{+0.44}_{-0.36}$   & $0.07^{+0.04}_{-0.05}$    & $0.56^{+0.05}_{-0.11}$    & $10.11^{+0.01}_{-0.13}$   \\
\hline
\end{tabular}

\medskip

Results of broken-linear fits to $\log \avis$ as a function of \logmstar{} (see
Eqn.~\ref{eq:barsize1}; parameter $\alpha_{2} = \alpha_{1} + (\beta_{1} - \beta_{2}) M_{\rm brk}$).
Parameter uncertainties are based on 2000 rounds of bootstrap resampling.

\end{minipage}
\end{table*}

\begin{table}
\caption{Fits to Bar Size versus \re{} and $h$}
\label{tab:bar-size-fits-reh}
\begin{tabular}{lcc}
\hline
Predictor    & $\alpha$ & $\beta$ \\
  (1)        & (2)          & (3) \\
\hline
$\log \re$ & $0.03^{+0.03}_{-0.03}$    & $0.45^{+0.05}_{-0.05}$    \\
$\log h$   & $0.04^{+0.02}_{-0.02}$    & $0.76^{+0.06}_{-0.06}$    \\
\hline
\end{tabular}

\medskip

Results of linear fits to bar size $\log \avis$ as a function of \logre{} and $\log h$.
Parameter uncertainties are based on 2000 rounds of bootstrap resampling.

\end{table}

\begin{table*}
\begin{minipage}{126mm}
\caption{Fits to Bar Size versus Stellar Mass and Galaxy Size}
\label{tab:bar-size-fits-multi}
\renewcommand{\arraystretch}{1.5}
\begin{tabular}{lccccc}
\hline
Sample & $\alpha_{1}$ & $\beta_{1}$ & $\beta_{2}$ & $\beta$  & $\log \: (M_{\rm brk} / \Msun)$  \\
  (1)  & (2)          & (3)         & (4)         & (5)      & (6)  \\
\hline
$\logmstar + \logre$ & $-0.44^{+0.26}_{-0.37}$   & $0.05^{+0.04}_{-0.03}$    & $0.47^{+0.09}_{-0.09}$    & $0.34^{+0.04}_{-0.05}$    & $10.24^{+0.13}_{-0.05}$   \\
$\logmstar + \log h$ & $0.02^{+0.29}_{-0.27}$    & $0.00^{+0.03}_{-0.03}$    & $0.35^{+0.06}_{-0.08}$    & $0.61^{+0.06}_{-0.06}$    & $10.13^{+0.02}_{-0.05}$   \\
\hline
\end{tabular}

\medskip

Results of fits to $\log \avis$ as a function of both \logmstar{}
\textit{and} \logre{} or $\log h$. Parameter uncertainties are based on
2000 rounds of bootstrap resampling. (The $\alpha_{2}$ parameter 
in Eqn.~\ref{eq:barsize3} can be derived from the parameters listed here.)

\end{minipage}
\end{table*}

\subsection{Bar Size as a Function of Galaxy Size: Half-light Radius $R_{e}$
and Disc Scale Length $h$}\label{sec:bar-size-galaxy-size} 

A number of studies have suggested that bar size is proportional to
galaxy \textit{size}, whether the latter is measured in terms of
isophotal radius (e.g., $R_{25}$), half-light radius \re, or exponential
disc scale length ($h$) -- although there is also evidence that this
scaling may vary with, for example, Hubble type or stellar mass
\citep{laine02,erwin05b,menendez-delmestre07,marinova07,laurikainen07,
aguerri09,dg16a}. Since galaxy size clearly scales with galaxy mass (see, e.g.,
\citealt{shen03,munoz-mateos15,lange15}, as well as
Section~\ref{sec:predict-galaxy-size} and
Figure~\ref{fig:galaxy-size-vs-mass}), it is plausible that the
correlation between bar size and stellar mass found in the previous
subsection could be an indirect one that reflects a more fundamental
correlation between bar and galaxy size (or bar and disc size).

I consider two different (though related) measurements of galaxy
``size''. The first is the galaxy half-light or effective radius \re{},
based on the fits of single, elliptical S\'ersic functions to the \sfourg{}
3.6\micron{} images by \citet{salo15}. The second is the \textit{disc}
exponential scale length $h$ from the multi-component (usually disc +
bulge or disc + bar + bulge) 2D fits of the same study. Since bars form
out of, and can also reshape, galaxy discs, we can a priori expect bar
size to be more directly related to $h$ than to \re. We can also expect
some scatter from the fitting process, as not all barred galaxies have
well-defined single-exponential outer discs, and even 2D decompositions
that incorporate bars may be biased by lenses, rings, mismatches between
models of bars and real galaxies, and so forth.

\begin{figure*}
\begin{center}
\hspace*{-4.5mm}
\includegraphics[scale=0.91]{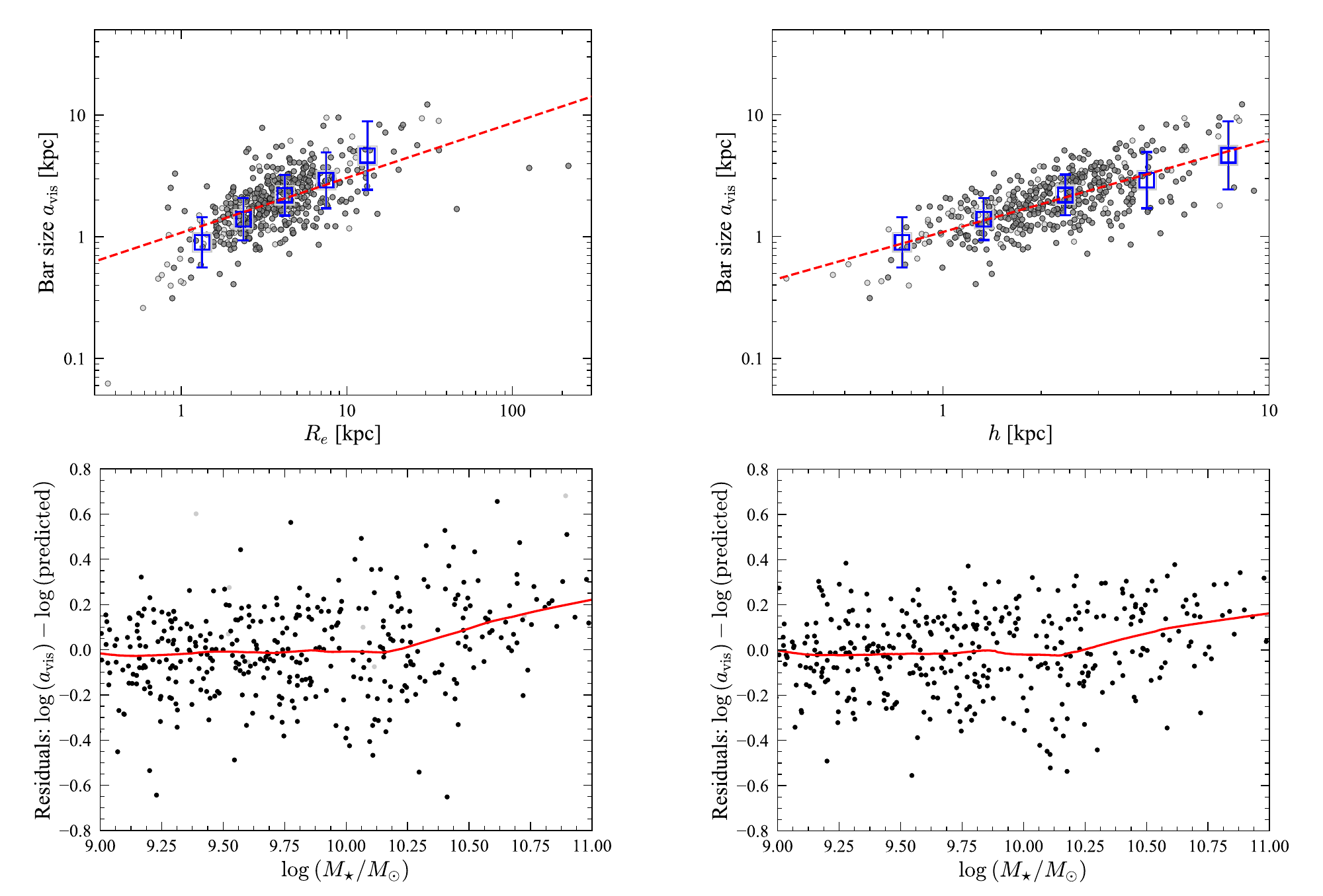}
\end{center}

\caption{Fits of bar size versus \re{} (left) and disc scale length $h$
(right), along with residuals from each fit plotted against stellar mass
(bottom two panels). Upper left: Bar size versus \re, with fit (dashed
red line). Dark grey circles are galaxies in the Main Barred
Spiral Sample, while light grey circles are additional barred \sfourg{}
galaxies with masses below or above the $\logmstar =$  9--11 limits of
the Main Spiral Sample; blue boxes and error bars indicate
logarithmically binned means and standard deviations for the Main Barred
Spiral Sample. Upper right: same, but now showing bar size versus disc
scale length $h$. Lower left: Residuals (Main Barred Spiral Sample) from
$\log \, \avis$--$\log \re$ fit plotted against \logmstar. Red line =
LOESS fit to residuals. Lower right: same as for lower left, except now
showing residuals from the $\log \, \avis$--$\log h$ fit. In both lower
panels, there is evidence for a broken-linear correlation between the
bar-size residuals and stellar mass, steepening for $\logmstar \ga 10.2$
-- similar to the broken-linear relation seen between bar size and
stellar mass (Figure~\ref{fig:mstar-fit}).
\label{fig:Re-h-fit-and-residuals}}
\end{figure*}

The upper panels of Figure~\ref{fig:Re-h-fit-and-residuals} show plots
of bar size \avis{} against galaxy half-light radius \re{} (upper left
panel) and disc scale length $h$ (upper right panel). Also plotted are
linear fits of $\log \avis$ as a function of $\log \re{}$ or $\log h$:
\begin{eqnarray}\label{eq:barsize2} 
\log \avis & = & \alpha + \beta \log \re , \\
\log \avis & = & \alpha + \beta \log h .
\end{eqnarray} 
The best-fit parameter values and uncertainties are listed in Table~\ref{tab:bar-size-fits-reh}.

The top panels of Figure~\ref{fig:Re-h-fit-and-residuals} show clear,
strong correlations between bar size and galaxy size; the correlation
appears to be tighter when disc scale length is used. (The
binned mean values suggest that the \avis-\re{} relation may become
nonlinear at high and low \re -- though the numbers become somewhat
sparse -- while the \avis-$h$ relation appears
consistently linear over the whole range of $h$.) Are these
correlations \textit{stronger} than the previously observed
bar-size--stellar-mass correlation? Table~\ref{tab:compare-fits}
compares the relative performance of the various fits in two ways.
First, assuming uniform 10\% errors on the bar sizes, I compute the
likelihood of the fits and then the Akaike Information Criterion (AIC)
values for each fit. Lower AIC values indicate relatively better fits,
with differences of $\Delta$AIC > 10 usually regarded as clear evidence
in favor of the fit with lower AIC.

Second, I estimate the mean squared prediction error (\msepred) for each
fit using bootstrap validation \citep[e.g.,][]{hastie09}. This means
generating a new dataset by bootstrap resampling from the original
dataset and then fitting the function to the bootstrapped data.
The data points which were \textit{not} included in the bootstrapped
data set are then used to compute residuals using the
bootstrap fit, storing the mean of the squared residuals from this
comparison. This is repeated 1000 times, and the mean of the
accumulated mean-squared residual values is computed; this is \msepred.
The smaller the \msepred, the better job a particular type of fit does
at predicting bar sizes.

As Table~\ref{tab:compare-fits} shows, the correlations between bar size
and galaxy size are better than the bar-size--\Mstar{} correlation, both
in terms of lower AIC values and lower \msepred{} values. (The same
basic result can be seen in Table~5 of \citealt{dg16a}, where the
Spearman correlation coefficient for different ranges of Hubble type is
always higher when \re{} or $h$ is the covariate than when \Mstar{} is.)
Furthermore, the bar-size--$h$ fit is clearly superior to the
bar-size--\re{} fit: bar size correlates better with the disc scale
length than with the galaxy half-light radius. Since bars originate from
disc instabilities, this is (as noted previously) to be expected.

\subsection{Bar Size as a Function of Multiple Parameters}\label{sec:bar-size-multiple} 

The previous section showed that the correlation between bar size
and galaxy size (either \re{} or $h$) is stronger than the correlation
with stellar mass. Even when the $\logavis$--$\logmstarshort$
relation is fit with a broken-linear function, the linear fits for $\logavis$
as a function of \logre{} or (especially) $\log h$ are better, with
smaller predicted residuals (upper part of Table~\ref{tab:compare-fits}).

So is the correlation between bar size and stellar mass merely a side
effect of the bar size--galaxy size correlation? After all, galaxy size
has long been known to correlate with galaxy luminosity and mass. In
fact, the galaxy size-mass relations in
Figure~\ref{fig:galaxy-size-vs-mass} show a (weak) break at $\logmstar
\sim 10.2$, with steeper slopes at larger stellar masses for both \re{}
and $h$. So it would seem possible that the real correlation is
therefore just that between bar size and galaxy size, with mass playing
no special role.

But this turns out to be only partly true. If we plot the
\textit{residuals} for the bar-size--galaxy-size fits against stellar
mass (lower panels of Figure~\ref{fig:Re-h-fit-and-residuals}), we can see
clear broken-linear trends, with the break at $\logmstarshort \sim
10.1$--10.2, very similar to the basic bar-size--\Mstar{} trend
(Figure~\ref{fig:mstar-fit}). This suggests that the dependence of
bar size on galaxy stellar mass is actually due to \textit{two} factors:
the linear\footnote{Linear, that is, in \textit{log} space.} relation
between bar size and galaxy size (combined with the bimodal
galaxy-size--\Mstar{} relation) \textit{and} an extra dependence of bar
size on stellar mass.

This motivates looking at bar size as a function of multiple parameters.
I do this by fitting the logarithm of bar size as a function of stellar
mass (using the broken-linear dependence that was successful in
Section~\ref{sec:barsize-mass}) and either \re{} or $h$:
\begin{equation}\label{eq:barsize3}
  \begin{aligned}
\log \avis & = & \alpha_{1} + \beta X + \beta_{1} \logmstarshort \;\; \mathrm{if} \; \logmstarshort < M_{\rm brk}, \\ 
& = & \alpha_{2} + \beta X + \beta_{2} \logmstarshort \;\; \mathrm{if} \; \logmstarshort > M_{\rm brk}
  \end{aligned}
\end{equation} 
where $X = \logre$ or $\log h$. (As in the case of Eqn.~\ref{eq:barsize1}, one of
the parameters can be expressed in terms of the others; e.g., 
$\alpha_{2} = \alpha_{1} + (\beta_{1} - \beta_{2}) M_{\rm brk}$). The
best-fit parameters are listed in Table~\ref{tab:bar-size-fits-multi},
and the fits are compared with the previous, single-variable fits in
Table~\ref{tab:compare-fits}. The AIC values in the latter table show that the
multi-variable fits are better than any of the single-variable fits: the
fit using both \re{} and stellar mass has $\Delta$AIC $\sim -1100$
relative to the fit using just \re, while the fit using both $h$ and
stellar mass has $\Delta$AIC $\approx -640$ relative to the fit using
just $h$. The MSE values are also lower for the multi-variable fits in
both cases.

As suggested by the pattern of residuals in the bottom panels of
Figure~\ref{fig:Re-h-fit-and-residuals}, which cluster about zero for
$\logmstar \la 10.1 \sim 10.2$, the stellar-mass dependence is indeed
bimodal, and in fact the slope of the mass dependence for $\Mstar <
M_{\rm brk}$ is basically indistinguishable from zero (in
Table~\ref{tab:bar-size-fits-multi}, $\beta_{1} = 0.05^{+0.04}_{-0.03}$
for the fit with \re, $\beta_{1} = 0.0 \pm 0.03$ for the fit with $h$).
This indicates that for galaxies with $\Mstar < M_{\rm brk}$, the bar
size really does depend on galaxy size (\re{} or $h$) alone; only for
more massive galaxies does an additional dependence on mass appear.

The best fit is clearly the one using both \Mstar{} and $h$: it has by far
the lowest AIC value ($\Delta$AIC $\approx 620$ less than the next-best
case) and the lowest MSE.  Given the superiority of the single-predictor
fit using $h$ compared to that using \re, this result is not surprising.
I do note that the fit using \Mstar{} and \re{} may often be easier to
apply than the \Mstar--$h$ fit, since \re{} values can be obtained by
simple S\'ersic fits to galaxy profiles or images (or by curve-of-growth
methods), while disc scale lengths require careful multi-component
decompositions and may be more difficult to achieve for small,
low-resolution images.

\begin{table}
\caption{Comparison of Fits for Bar Size}
\label{tab:compare-fits}
\begin{tabular}{lrr}
\hline
Predictor(s)      & AIC  & \msepred{} \\
\hline
\logmstar{} (linear)            & 8520.0  & 0.046 \\
\logmstar{} (broken-linear)     & 7870.1  & 0.042 \\
$\log \, \re$                   & 7328.8  & 0.039 \\
$\log \, h$                     & 6242.2  & 0.033 \\
$\log \, \fgas$                 & 10794.1 & 0.058 \\
\hline
$\log \, \re \; + \; \logmstar$ (broken-linear) & 6203.6 & 0.034 \\
$\log \, h \; + \; \logmstar$ (broken-linear)   & 5583.5 & 0.030 \\
\end{tabular}

\medskip

Comparison of fits of different models for logarithmic bar size $\log
\avis$ in kpc. Fits were done to barred galaxies in the Main Spiral
Sample (galaxies with $\logmstar = 9$--11 and valid \re{} and $h$
values). (1) Predictor variable(s). (2) Corrected Akaike Information
Criterion value for fit; lower values indicate better fits. (3) Mean
squared prediction error for log of bar size (kpc), based on 1000 rounds
of bootstrap validation.

\end{table}

\section{Possible Correlations with Other Parameters}\label{sec:other-params} 

\subsection{Gas Mass Fraction}\label{sec:gas-fraction} 

A number of theoretical studies have suggested that bar formation and
growth may be retarded by a sufficiently high gas mass fraction
\citep[e.g.,][]{berentzen98,
berentzen07,bournaud05c,villa-vargas10,athanassoula13}. The models
suggest that this should generally be an \textit{anti}-correlation, with
higher gas mass fractions associated with \textit{shorter} bars.

Whether the gas mass fraction observed \textit{at the present
day} correlates well -- or at all -- with the fraction during bar
formation and growth is unclear, since gas can be consumed (and also
expelled) by star formation, stripped by tidal interactions or ram
pressure, etc., which would convert a gas-rich galaxy into a gas-poor
galaxy. Nonetheless, it is possible that a weak correlation could still
exist at $z = 0$; the simulations of \citet{athanassoula13}, which included star
formation, seem to suggest that the $z = 0$ gas mass fractions are
correlated with their initial fractions (see the discussion in
Section~\ref{sec:discuss-gas}).

\begin{figure}
\begin{center}
\hspace*{-4mm}
\includegraphics[scale=0.47]{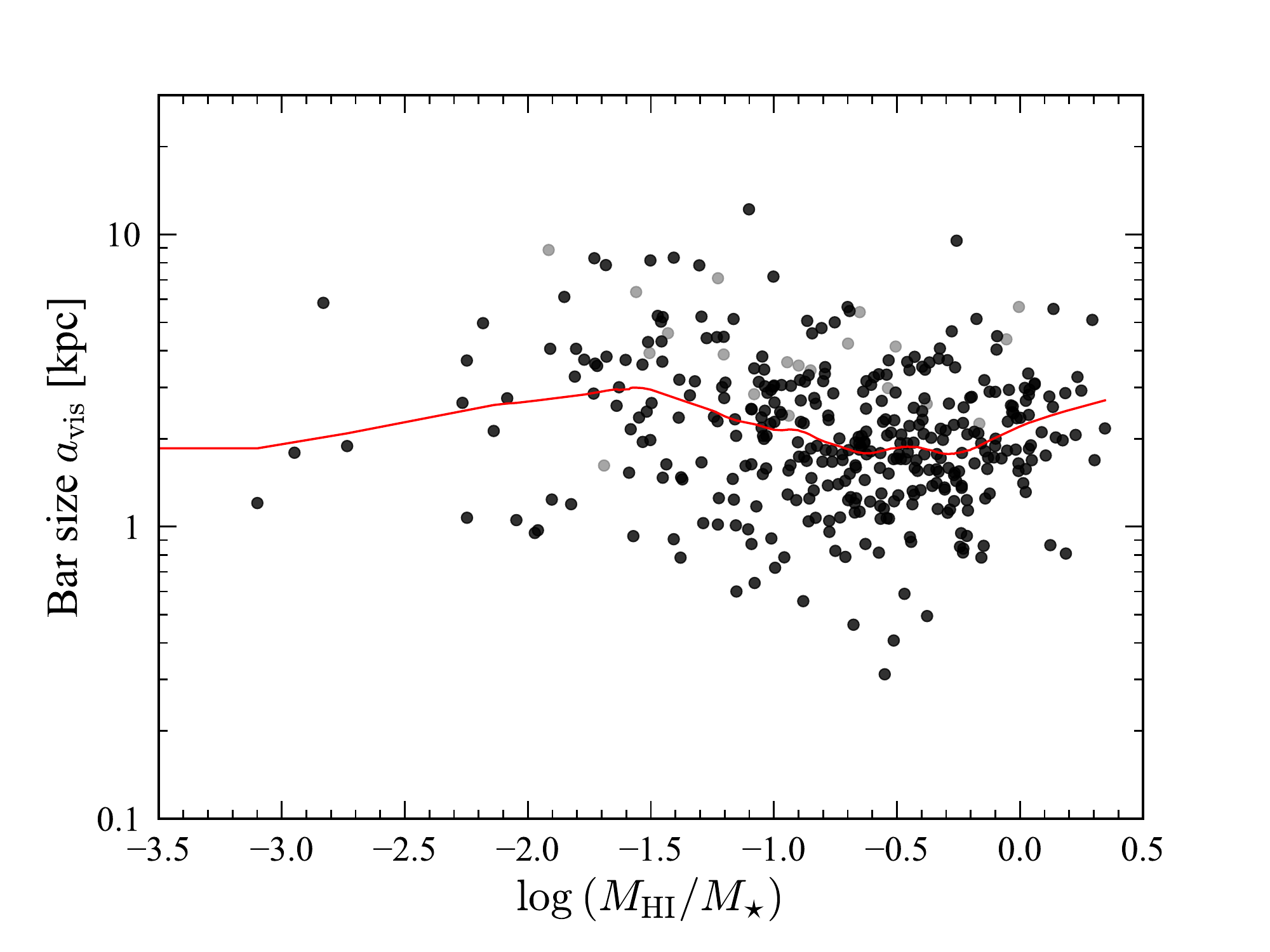}
\end{center}

\caption{Bar size as a function of gas mass fraction for the Parent Barred Spiral
Sample (grey and black points) and the Main Barred Spiral Sample (black points).
The red line indicates a LOESS fits to the individual points in the Parent Barred
Spiral Sample.
\label{fig:barsize-vs-fgas}}
\end{figure}

Figure~\ref{fig:barsize-vs-fgas} shows how bar size depends on gas mass
fraction in the \sfourg{} galaxies. There is at first glance a
suggestion of an overall trend in accord with the predictions: more
gas-rich galaxies do tend to have smaller bars.

The problem with this idea is that the gas mass fraction is rather
strongly anti-correlated with stellar mass (e.g.,
Figure~\ref{fig:app-fgas-vs-mstar}), so we could just be seeing a side
effect of the latter relation. The fact that the trend in
Figure~\ref{fig:barsize-vs-fgas} is clearly weaker than the trends of
bar size with stellar mass (Figure~\ref{fig:mstar-fit}) and with galaxy
size (Figure~\ref{fig:Re-h-fit-and-residuals}) -- and that the trend is
possibly not even monotonic -- suggests that this might be the case.
Moreover, the fit of bar size versus \fgas{} (see
Table~\ref{tab:compare-fits}) is clearly worse than all of the other
fits in terms of AIC, which makes it less likely that \fgas{} is a
key factor in explaining bar size.

\begin{figure*}
\begin{center}
\hspace*{-2.5mm}
\includegraphics[scale=0.92]{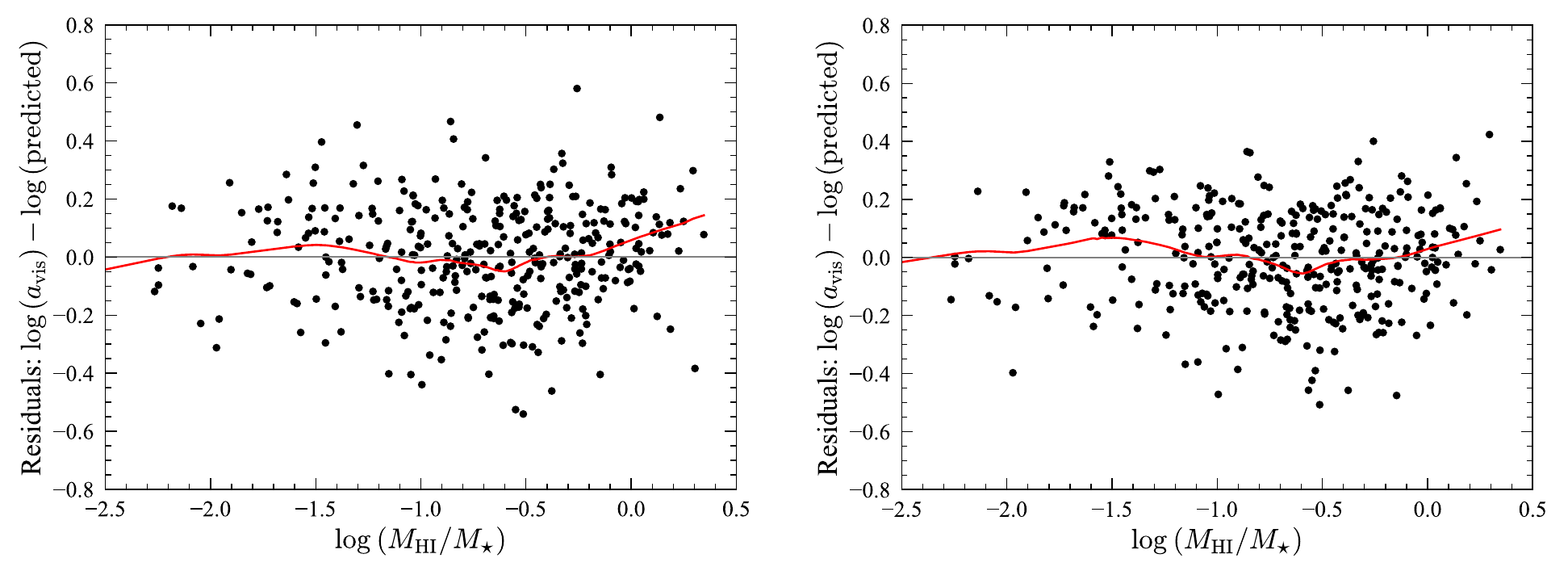}
\end{center}

\caption{Residuals from the fit of bar size as a function of both stellar mass
and \re{} (left panel) or stellar mass and disc scale length $h$ (right panel),
plotted against gas mass fraction.
Red lines indicate LOESS fits to the residual points. The absence of any strong
trends in the residuals indicates that gas mass fraction is \textit{not} a
meaningful predictor of bar size.
\label{fig:resid-vs-fgas}}
\end{figure*}

It is still possible that gas fraction could explain \textit{some} of
the variance in the other fits -- i.e., gas fraction might correlate
with deviations from the bar-size--stellar-mass or bar-size--galaxy-size
fits. Figure~\ref{fig:resid-vs-fgas} explores this by showing, as a
function of gas fraction, the residuals from the fits of bar size as a
function of stellar mass and galaxy size (either \re{} or $h$) from
Table~\ref{tab:bar-size-fits-multi}. There is very little correlation:
galaxies with bars which are larger or smaller than what is typical for
their stellar mass and size are \textit{not} systematically gas-rich
\textit{or} gas-poor (Spearman correlation coefficient $r = 0.08$, $P = 0.11$).
There is \textit{perhaps} a weak turn-up for very gas-rich galaxies
($\logfgas > -0.5$), but this is actually in the opposite sense from
what models have suggested: very gas-rich galaxies tend to have slightly
\textit{larger} bars.

A potential issue is the fact that I have only considered
\textit{atomic} gas content (from \hi{} observations) for the \sfourg{}
spirals; molecular gas is not being counted. However, observations suggest
that molecular gas is typically only $\sim 25$--30\% of the atomic gas
mass in $\logmstar = 9$--11 spirals, and is more abundant relative to
atomic gas in \textit{higher-mass} galaxies \citep[e.g.,][]{boselli14b}.
This would actually make the disagreement with theory worse: the true
gas mass fractions should generally be larger in precisely those
galaxies which have larger bars.

I conclude that bar size has no significant relation to (present-day)
gas mass fraction.

\subsection{Hubble Type} 

\begin{figure}
\begin{center}
\hspace*{-4.5mm}
\includegraphics[scale=0.48]{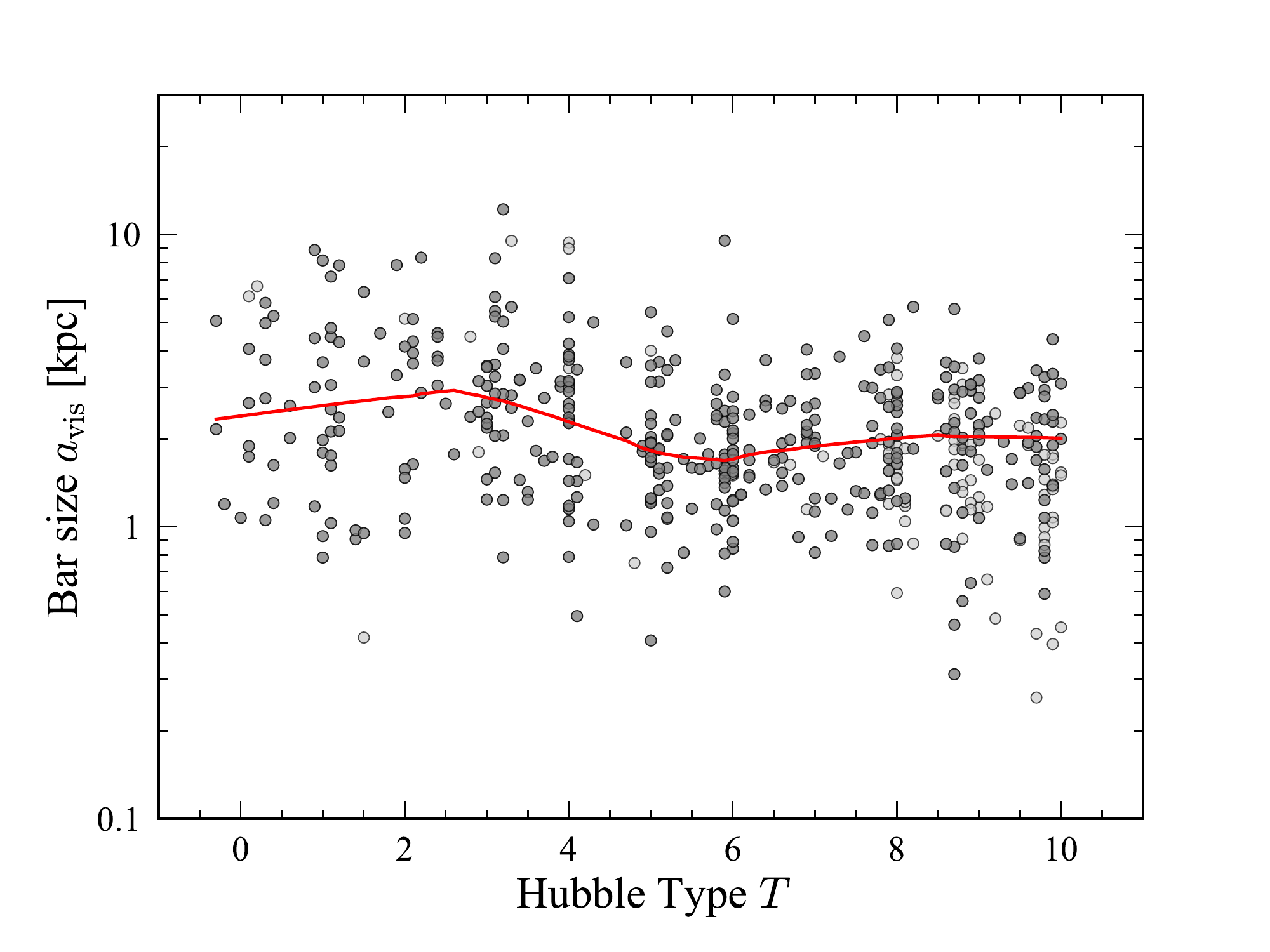}
\end{center}

\caption{Bar size as a function of Hubble type $T$ for \sfourg{} spiral
galaxies with $D \leq 30$ Mpc; galaxies in the Parent Spiral Sample are
indicated by darker data points.
The red line indicates a LOESS fit to the latter data points.
\label{fig:barsize-vs-T}}
\end{figure}

\begin{figure*}
\begin{center}
\hspace*{-2.5mm}
\includegraphics[scale=0.92]{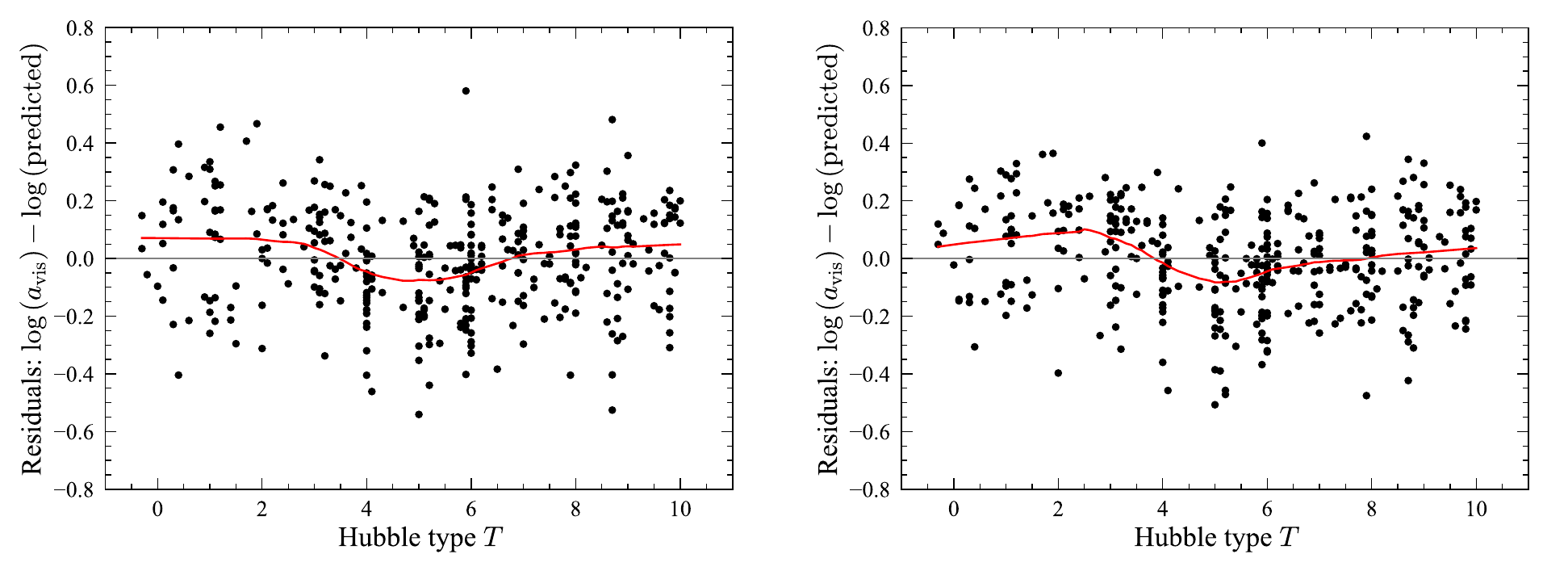}
\end{center}

\caption{Residuals from the fit of bar size as a function of both stellar mass
and \re{} (left panel) or stellar mass and disc scale length $h$ (right panel), plotted
against Hubble type $T$.
Red lines indicate LOESS fits to the residual points.
\label{fig:resid-vs-T}}
\end{figure*}

Most studies of bar sizes have tended to focus on trends with Hubble
type, either in very broad general terms -- ``early-type'' versus
``late-type'' disc galaxies -- or in finer detail
\citep[e.g.,][]{ee85,martin95,erwin05b,menendez-delmestre07,dg16a}. In
Erwin (2005), for example, I argued that bars in Hubble types Sc--Sd
were systematically smaller than bars in S0--Sab galaxies, by a factor
of roughly two.

But comparison of Figure~\ref{fig:barsize-vs-T} -- which shows \sfourg{}
bar size versus Hubble type\footnote{Hubble types are taken from the \sfourg{}
main catalog and are ultimately from the HyperLEDA database (http://leda.univ‐lyon1.fr/).} -- with the trends of bar size versus
stellar mass or (especially) versus galaxy size
(Figures~\ref{fig:mstar-fit} and \ref{fig:Re-h-fit-and-residuals}) shows
that there the trend of bar size with Hubble
type is quite weak.\footnote{For example, the Spearman correlation coefficient for
barred galaxies in the Main Spiral Sample is 0.56 ($P = 3 \times
10^{-32}$) for bar size versus stellar mass and 0.61 ($P = 4 \times
10^{-39}$) for bar size versus disc scale length, but only $-0.18$ ($P =
7 \times 10^{-4}$) for bar size versus Hubble type.}

This trend \textit{does} reproduce the
known tendency for early-type spirals to have somewhat larger bars than
later-type spirals (mean $\avis \sim 2.5$ kpc for S0/a--Sb versus $\sim
1.5$ kpc for Sc and later). But it is quite possible that this trend
could simply be a side effect of the strong correlations between Hubble
type and stellar mass (and between Hubble type and galaxy size):
late-type spirals tend to be lower mass and smaller. The question then
becomes: are there any residual trends of bar size with Hubble type
\textit{once the stellar-mass and galaxy-size dependence of $T$ is taken
care of?}

Figure~\ref{fig:resid-vs-T} plots residuals from the fits of bar size to
the combination of stellar mass and galaxy size (\re{} or $h$) versus
Hubble type $T$. For galaxies with $\logmstar = 9$--11, the Spearman
correlation coefficients for the bar-size residuals are $-0.09$ and $-0.009$
($P = 0.094$ and 0.87) for the $\re$-based and $h$-based fits,
respectively. There is thus no evidence for a systematic
dependence of bar size on Hubble type, once the underlying correlation
with stellar mass is accounted for (apart from some evidence for a
minimum in bar size for $T \sim 5$).

\section{Connections Between Bar Presence, Bar Size, and Galaxy Size: Using
Bars to Predict Galaxy Sizes}\label{sec:predict-galaxy-size} 

\begin{figure*}
\begin{center}
\hspace*{-14mm}\includegraphics[scale=0.55]{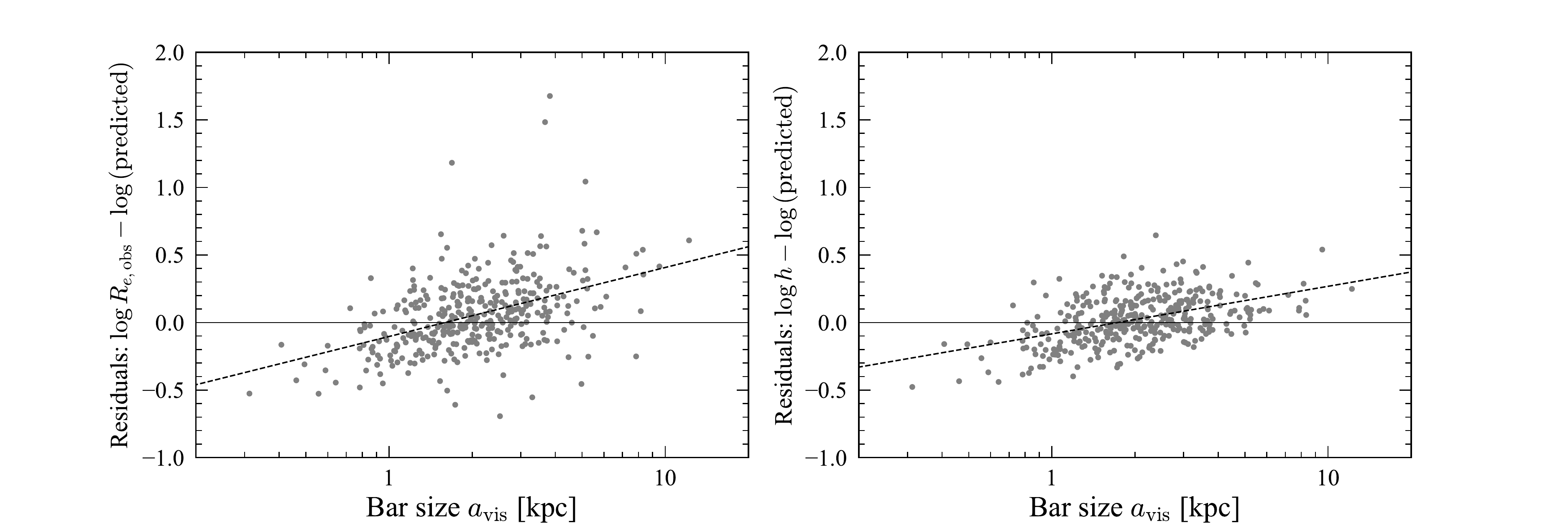}
\end{center}

\caption{Residuals of the galaxy-size--stellar-mass fit for $R_{e}$ (left) and
for disc scale length $h$ (right) as a function of bar size. (See Appendix~\ref{app}
for details of these fits.) The dashed lines are simple linear fits to the residual
data points. In both cases, there is a correlation: galaxies which are larger
relative to the typical size for their stellar mass also have larger bar sizes.
\label{fig:galaxy-size-residuals-vs-bar-size}}
\end{figure*}

\begin{figure*}
\begin{center}
\hspace*{-2mm}
\includegraphics[scale=0.56]{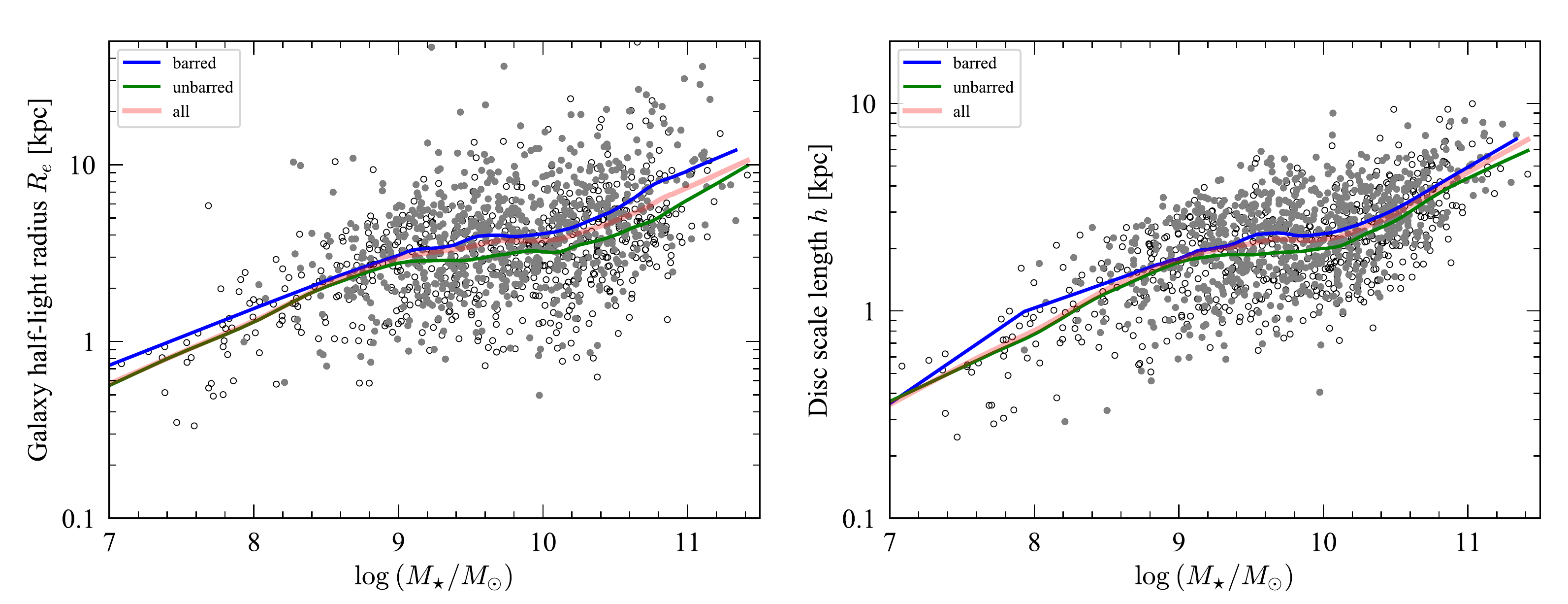}
\end{center}

\caption{Galaxy size $R_{e}$ (left) or disc scale length $h$ (right) as
a function of stellar mass for \sfourg{} galaxies. Filled symbols are
barred galaxies and open symbols are unbarred. The three lines represent
LOESS fits to just the barred galaxies (blue line), just the unbarred
galaxies (green line), and to all the galaxies (light red line). All
galaxies, barred and unbarred, show the same general trend; nonetheless,
galaxies with bars are, on average, larger than unbarred galaxies of the
same mass over the entire range. \label{fig:galaxy-size-vs-mass}}
\end{figure*}

In Sections~\ref{sec:barsizes} and \ref{sec:other-params}, I looked at
the degree to which global galaxy properties -- total stellar mass,
size, gas content, Hubble type -- could explain (or, more properly,
predict) the sizes of bars in barred galaxies. The strongest individual
relation is clearly that between galaxy size and bar size.

This relation implies a trivial inversion: we can, in principle, use bar
size to predict galaxy size. Evidence for this is shown in
Figure~\ref{fig:galaxy-size-residuals-vs-bar-size}, which plots
residuals of galaxy size (\re{} or $h$) from broken-linear fits to the
\re--stellar-mass and $h$--stellar-mass relations (see
Appendix~\ref{app}) versus \textit{bar} size. For both \re{} and $h$
there is a clear trend: galaxies with larger bars have larger half-light
and disc sizes, even when stellar mass is accounted for.

A related implication is the possibility that galaxy size might also
depend on the mere \textit{presence} of bars. Are galaxies with bars
systematically larger (or smaller) than galaxies without bars? Because
simulations of bar formation and growth often show accompanying changes
to the stellar distribution, it is possible to argue that part of the
connection between bar size and galaxy size is due to bars driving
changes in galaxy size, rather than the other way around.

There is already evidence that the presence or absence of bars does
correlate with galaxy size, in the sense that a galaxy with a bar is
more extended than a galaxy of the same mass without a bar.
\citet{sanchez-janssen13} found, for a sample of massive, face-on SDSS
galaxies with 2D bulge/bar/disc decompositions, that barred galaxies had
disc scale lengths $\sim 13\%$ larger than unbarred galaxies of the same
mass. More recently, \citet{dg16b} presented plots of stacked galaxy
stellar surface-mass-density profiles, derived from \sfourg{} data,
divided into different mass and Hubble-type bins, and also into barred
versus unbarred subsets. At least for galaxies with $\logmstarshort >
9$, the profiles of barred and unbarred galaxies with similar masses
differ (see their Figure 5 and 6): barred galaxies have disc profiles
which are shallower and more extended than unbarred galaxies with
similar masses.

We can see this effect in the global $R_{e}$ and disc size measurements
for the \sfourg{} galaxies. Figure~\ref{fig:galaxy-size-vs-mass} shows
how galaxy size ($R_{e}$ or $h$) depends on stellar mass for the main
\sfourg{} spiral sample considered in this paper. Also plotted are LOESS
fits to the data for all galaxies (light red line), for barred galaxies
(blue line), and for unbarred galaxies (green line). The fit lines show
that barred galaxies are, on average, slightly larger than unbarred
galaxies of the same stellar mass, and that this is true over most of
the mass range -- certainly for $\logmstar \ga 9$ (at smaller masses,
the number of barred galaxies drops off and the barred-galaxy fit line
becomes unreliable).

Taken together, this suggests that galaxy size can be considered a function of both
bar size and bar \textit{presence}: galaxies with bars tend to be
somewhat larger than unbarred galaxies, and galaxies with big bars
tend to be larger than galaxies with small bars. 
We can express this general dependence in simple form as follows:
\begin{eqnarray}\label{eq:galaxy-size} 
\logre  & = &  f_{\rm bl}(\logmstarshort) + B (\alpha + \beta \logavis) \\
\log h  & = &  f_{\rm bl}(\logmstarshort) + B (\alpha + \beta \logavis)
\end{eqnarray} 
where $f_{\rm bl}$ is the broken-linear dependence modeled in
Section~\ref{sec:bar-size-galaxy-size} and $B$ is a binary variable, which is 0 for unbarred
galaxies and 1 for barred galaxies. The results of fitting these
relations to the data are shown in Table~\ref{tab:fits-galaxy-size}. For both $\re$ and $h$,
including bar presence/size leads to significantly better fits (e.g.,
$\Delta$AIC $\sim -2500$ to $-3900$).

\begin{table*}
\begin{minipage}{\linewidth}
\caption{Fits to Galaxy Size Versus Stellar Mass and Bar Size}
\label{tab:fits-galaxy-size}
\begin{tabular}{lccccccccc}
\hline
Predictor(s) & $\alpha_{1}$ & $\beta_{1}$ & $\beta_{2}$ & $\log \: (M_{\rm brk} / \Msun)$  & $\alpha$ & $\beta$ & AIC & MSE$_{\rm pred}$\\
  (1)        & (2)          & (3)         & (4)         & (5)                              & (6)      & (7)     & (8) & (9) \\
\hline
\multicolumn{9}{c}{Relation for $\logre$} \\
\hline
$\logmstar$          & $-0.10^{+0.34}_{-0.40}$   & $0.06^{+0.04}_{-0.04}$    & $0.38^{+0.08}_{-0.08}$    & $10.08^{+0.12}_{-0.12}$   & \ldots & \ldots & 22930.8 & 0.077 \\
$\logmstar + \log \avis$ & $0.31^{+0.32}_{-0.49}$    & $0.01^{+0.05}_{-0.03}$    & $0.26^{+0.06}_{-0.08}$    & $10.00^{+0.16}_{-0.11}$   & $-0.04^{+0.02}_{-0.03}$   & $0.58^{+0.06}_{-0.06}$    & 18426.2 & 0.062 \\
\hline
\multicolumn{9}{c}{Relation for $\log h$} \\
\hline
$\logmstar$          & $-0.25^{+0.40}_{-0.22}$   & $0.06^{+0.02}_{-0.04}$    & $0.41^{+0.04}_{-0.11}$    & $10.09^{+0.05}_{-0.19}$   & \ldots & \ldots & 10038.8 & 0.034 \\
$\logmstar + \log \avis$ & $-0.12^{+0.30}_{-0.27}$   & $0.04^{+0.03}_{-0.03}$    & $0.31^{+0.04}_{-0.14}$    & $10.09^{+0.12}_{-0.25}$   & $-0.07^{+0.02}_{-0.02}$   & $0.40^{+0.03}_{-0.04}$    & 8258.2 & 0.028 \\
\hline
\end{tabular}

\medskip

Fits of logarithm of galaxy half-light size \re{} or disc scale length
$h$ (kpc) for all galaxies in the Main Spiral Sample (both barred and
unbarred) as a function of stellar mass and (optionally) bar size. The
first fit in each case is a broken-linear function of \logmstar{} (as in
Eqn.~\ref{eq:barsize1}); the second fit includes an additional linear
function of bar size (see Eqn.~\ref{eq:galaxy-size}). (1) Type of fit.
(2)--(7) Best fit parameter values and uncertainties (from 2000 rounds
of bootstrap resampling). (8) Corrected Akaike Information Criterion
value for fit (smaller values indicate better fits). (9) Mean squared
prediction error for log of galaxy size (\re{} or $h$, kpc), based on
1000 rounds of bootstrap validation.

\end{minipage}
\end{table*}

\section{Discussion}\label{sec:discuss} 

\subsection{The Size and Presence of Bars Does Not Depend on (Present-Day) Gas Fraction}\label{sec:discuss-gas} 

A number of theoretical studies have suggested that a high gas mass
fraction in disc galaxies can delay bar formation and also limit secular
bar growth after formation. For example, \citet{villa-vargas10} found
that final bar sizes were more than 50\% smaller in simulations with
(initial) gas fractions of $> 10$\% versus those with gas fractions $<
5$\%; they did note that their simulations did not include star
formation, which might in principle alter these results. The simulations
of \citet{athanassoula13} looked at the effects of halo shape and
initial gas fraction (ranging from 0 to 100\%) on bar formation and
growth, for model galaxies with a fixed baryonic mass of $5 \times
10^{10} \Msun$. These simulations \textit{did} include star formation,
and the final gas fractions were in fact roughly an order of magnitude
smaller than the initial fractions. Nonetheless, the final fractions
were correlated with the initial values (e.g., their Fig.~2 and
Table~2). Although the focus in that study was on bar \textit{strength}
rather than size, their Figs.~4 and 5 clearly indicate that simulations
with higher initial gas fractions tend to end up with shorter final bar
sizes, so there should be a correlation between bar size and $z \sim 0$
gas fraction.

But observationally (Section~\ref{sec:gas-fraction}), there is
effectively no connection between the present-day gas fraction and bar
size. In \citetalias{erwin18} (Section~6.1), I showed that the mere
\textit{presence} of bars in \sfourg{} does not depend on present-day
gas fraction either, even when the strong dependence of bar fraction on
stellar mass is accounted for. If, as simulations suggest, gas fraction
\textit{does} significantly affect bar formation and growth, then this
relation must somehow be completely absent by $z = 0$.

\subsection{Which Way(s) Does the Arrow of Causality Point?} 

As shown previously, the best single-parameter correlation is that
between bar size and disc scale length, and the best multi-parameter
correlation for bar size is with the combination of stellar mass and
disc scale length. 

At first glance, this suggests that the bar size--disc size correlation
is the most fundamental one. One can easily imagine a direct causal
relation between disc size and bar size: bars are disc phenomena,
forming out of the disc, and thus larger discs naturally give rise to
larger bars. 

This may not be the whole story, however. Numerous simulations have
found that bar formation and growth redistribute stars (and gas) in the
disc, typically leading to an increase in the disc scale length 
\citep[e.g.,][]{hohl71,valenzuela03,debattista06}.
This raises the possibility that bar and disc sizes are coupled by a
feedback process, so that the growth in bar size is potentially one of the
\textit{causes} of disc size growth. The fact that galaxy size is partly
dependent on the mere \textit{presence} of a bar
(\citealt{sanchez-janssen13}; Section~\ref{sec:predict-galaxy-size}) --
that is, galaxies with bars have, on average, more extended discs -- is
consistent with this idea.

Nonetheless, it is important to note that the broken-power-law
correlation between galaxy size (\re{} or $h$) and stellar mass exists
for \textit{unbarred} galaxies as well as for barred galaxies (e.g.,
Figure~\ref{fig:galaxy-size-vs-mass}). This seems to be true even for
the poorly sampled low-mass end (e.g., $\logmstar < 9$ in
Figure~\ref{fig:galaxy-size-vs-mass}), where the bar fraction becomes
quite low. Unless the majority of today's unbarred galaxies were
previously barred, this implies that the fundamental size-mass relation
of galaxies is not driven by bar formation or evolution.

This leads me to suggest that a possible causal hierarchy might be
something like the following: 
\begin{enumerate}

\item Some general process of galaxy formation and evolution sets the
trends of galaxy (and especially disc) size as a function of stellar
mass, as seen in Figure~\ref{fig:galaxy-size-vs-mass}. When bars form,
they inherit their size primarily from the disc size, resulting in (over
the mass range of $\logmstar = 9$--11, at least) the bar-size--stellar-mass
correlation seen in Figure~\ref{fig:mstar-fit}.

\item The formation and growth of bars causes, in turn, an increase in
galaxy size, one which is (at least approximately) proportional to the
growth in bar size.

\end{enumerate}

We are, admittedly, still left with the puzzling fact that bar size
apparently \textit{also} correlates with stellar mass to a certain
degree \textit{independently of} its correlation with disc size
(Section~\ref{sec:bar-size-multiple}), something that cannot be
accounted for with the preceding argument.

\subsection{So, What \textit{Does} Determine Bar Size?}

One possibility is that we are seeing a temporal effect: more massive
galaxies formed cool disks earlier, so they became bar-unstable and
formed their bars earlier \citep[as suggested by,
e.g.][]{sheth12,kraljic12}, and thus they have had more time for their
bars to grow in length. Observations using \textit{HST} have generally
indicated that the frequency of bars at higher redshifts is larger in
the most massive galaxies. In this scenario, the highest-mass galaxies
in the local Universe would tend to have unusually large bars because
their bars have had the most time to grow in size. The main problem (in
addition to the suspicion that galaxy formation might not be coordinated
enough in time across all environments) is that it's unclear why this
should only be true for galaxies with $\logmstar \sim 10.1$ or greater
-- why have all the bars in lower-mass galaxies not had any time to grow
in size? It is somewhat hard to believe that this is because they have
all formed so recently that they haven't had time to grow; after all,
the $z \sim 0$ bar fraction is highest for $\logmstar \sim 9.7$
\citep{erwin18}.

\section{Summary}\label{sec:summary} 

This paper has presented an analysis of how the sizes (deprojected semi-major
axis \avis) of bars depend -- or do \textit{not} depend -- on their host
galaxy properties: specifically, how bar size relates to stellar mass,
galaxy size (half-light radius \re{} or exponential-disc scale length
$h$), atomic gas content, and Hubble type.

A strong correlation exists between bar size and galaxy stellar mass, as
has long been noted (usually using galaxy luminosity rather than stellar
mass). However, this correlation is actually \textit{bimodal}, with a
very shallow relation for low-mass galaxies ($\avis \propto
\Mstar^{0.1}$) which steepens rather abruptly ($\avis \propto
\Mstar^{0.6}$) for stellar masses with $\logmstar \ga 10.1$.

Bar size also correlates strongly with galaxy size, with the strongest
correlation being with disc scale length: $\avis \propto \re^{0.5}$,
$\avis \propto h^{0.8}$. Both correlations are stronger than the
correlation with galaxy mass. However, even though the
bar-size--galaxy-size correlation can partly explain the
bar-size--galaxy-mass correlation (because galaxy size itself correlates
with stellar mass, with more massive galaxies being more extended),
there is still a residual correlation with stellar mass itself. 

More precisely, for lower-mass galaxies ($\logmstar \la 10.2$), bar size
depends on galaxy size only. For higher-mass galaxies, stellar mass also
matters: given two galaxies with the same size, the galaxy with higher
mass will tend to have a longer bar. Thus, a better general predictor of
bar size is the \textit{combination} of galaxy size and stellar mass.

Once the correlations with galaxy size (\re{} or $h$) and stellar mass
are accounted for, bar size shows essentially no residual correlation
with either atomic gas mass fraction ($\MHI/\Mstar$) or Hubble type. The
first result is perhaps in conflict with theoretical arguments
suggesting that high gas fractions can either delay bar formation or
slow the growth of bar sizes, or both, although it is unclear how
well or poorly present-day gas content should correlate with prior gas content
(i.e., during the time of bar formation and growth). The second result
indicates that classic arguments for different bar sizes as a function
of Hubble type \citep[e.g.,][]{ee85,martin95,erwin05b} appear to be side
effects of the general tendency of later Hubble types to be smaller and
lower in mass than early-type spirals.

Finally, I note that bars can be used as predictors of galaxy size. This
is not only true for barred galaxies -- at constant mass, galaxies with
larger bars have, on average, larger \re{} and $h$ -- but also reflects
the fact that barred galaxies as a class tend to be more extended than
unbarred galaxies of the same mass.

\section*{Acknowledgments} 

I would like to thank Victor Debattista for numerous helpful discussions
at various stages of this project, Martin Herrera-Endoqui for answering
questions about the \sfourg{} measurements, Adriana de
Lorenzo-C{\'a}ceres and Jairo M{\'e}ndez Abreu for careful reading of an
earlier version, and the referee for various interesting comments
and suggestions.

This work is based in part on observations made with the
\textit{Spitzer} Space Telescope, obtained from the NASA/IPAC Infrared
Science Archive, both of which are operated by the Jet Propulsion
Laboratory, California Institute of Technology under a contract with the
National Aeronautics and Space Administration. This paper also makes use of
data obtained from the Isaac Newton Group Archive which is maintained as
part of the CASU Astronomical Data Centre at the Institute of Astronomy,
Cambridge.

This research also made use of Astropy, a community-developed core Python
package for Astronomy \citep{astropy13}.

\bibliographystyle{mnras}

\appendix{}

\section{Galaxy Scaling Relations}\label{app}

\begin{figure*}
\begin{center}
\hspace*{-17mm}\includegraphics[scale=0.56]{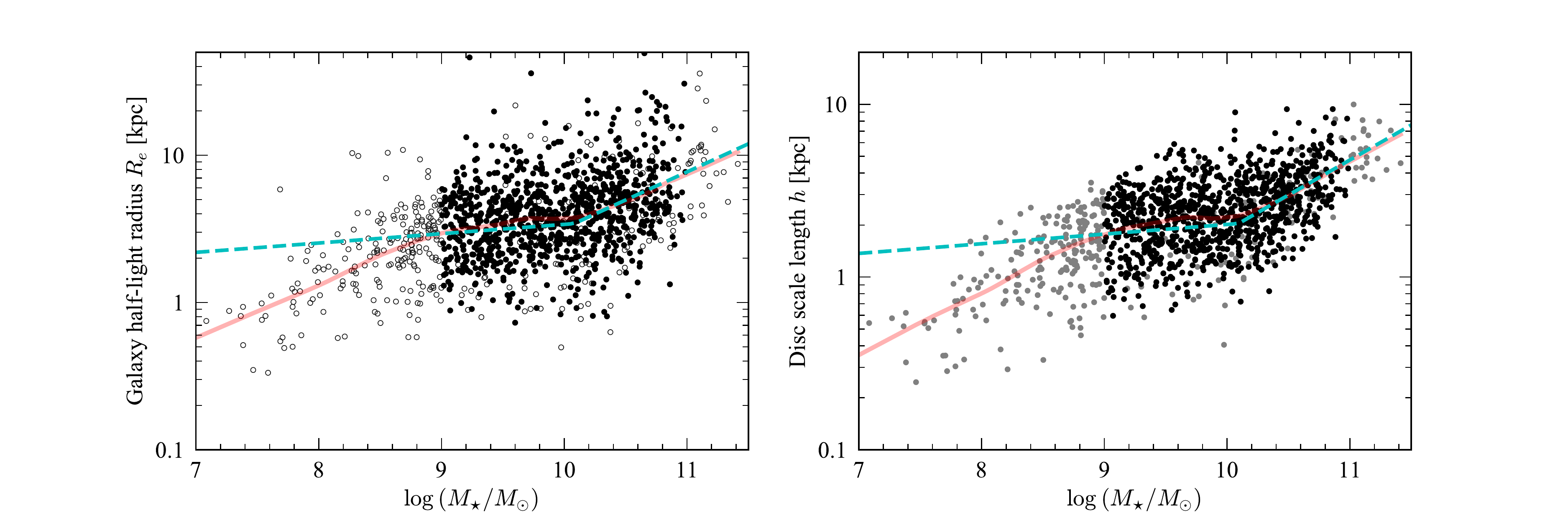}
\end{center}

\caption{Galaxy sizes as a function of stellar mass for \sfourg{} spiral
galaxies. Left: Half-light radius \re{} (from single-S\'ersic fits in
\citealt{salo15}) versus stellar mass. Right: main disc exponential
scale length $h$ (from multi-component fits in \citealt{salo15}). Filled
black symbols are galaxies in the Main Spiral Sample ($D \leq 30$ Mpc,
$\logmstar = $9--11); open symbols are other \sfourg{} galaxies. Red
lines indicate LOESS fits to the $D \leq 30$ Mpc data; dashed cyan lines
are broken-linear fits to the Main Spiral Sample galaxies.
\label{fig:app-galaxysize-vs-mstar-with-fits}}
\end{figure*}

In Figure~\ref{fig:app-galaxysize-vs-mstar-with-fits} I show the general
size-mass scaling relations for the entire \sfourg{} spiral sample (grey
symbols); LOESS fits to the $D \leq 30$ Mpc subsample are shown with red
lines. Galaxies in the Main Spiral Sample are plotted using black
symbols. For both \re{} and $h$, there is a consistent trend: a steep
relation for low masses ($\logmstarshort \la 9$), a shallow trend for
higher masses, and then steeper again for $\logmstarshort \ga 10.1$.

Since the Main Spiral Sample is restricted to stellar masses of
$\logmstarshort = 9$--11, I also show the result of a broken-linear fit
(in log space) to that data using a dashed blue line; these fits were
done in a fashion very similar to the broken-linear fits (for bar size)
in Section~\ref{sec:barsize-mass}. The best-fit coefficients are
$\alpha_{1} = -0.10$, $\beta_{1} = 0.063$, $\beta_{2} = 0.38$, and
$\log(M_{\rm brk}/\Msun = 10.08$ for the $\re$--\Mstar{} relation and
$\alpha_{1} = -0.25$, $\beta_{1} = 0.056$, $\beta_{2} = 0.41$, and
$\log(M_{\rm brk}/\Msun = 10.09$ for the $h$--\Mstar{} relation


The $\log \re$--$\logmstarshort$ trend appears to be roughly consistent
with local size-mass relationships as plotted for S4G in
\citet{munoz-mateos15}, along with trends for SDSS \citep{shen03} and
GAMA \citep{lange15}. The $\log h$--$\logmstarshort$ trend clearly has
less scatter. This may be due to the higher accuracy of 2D fits that include
multiple components, as opposed to forcing galaxies with multiple components
to be fit with a single S{\'e}rsic function.

\begin{figure}
\begin{center}
\hspace*{-3.5mm}\includegraphics[scale=0.47]{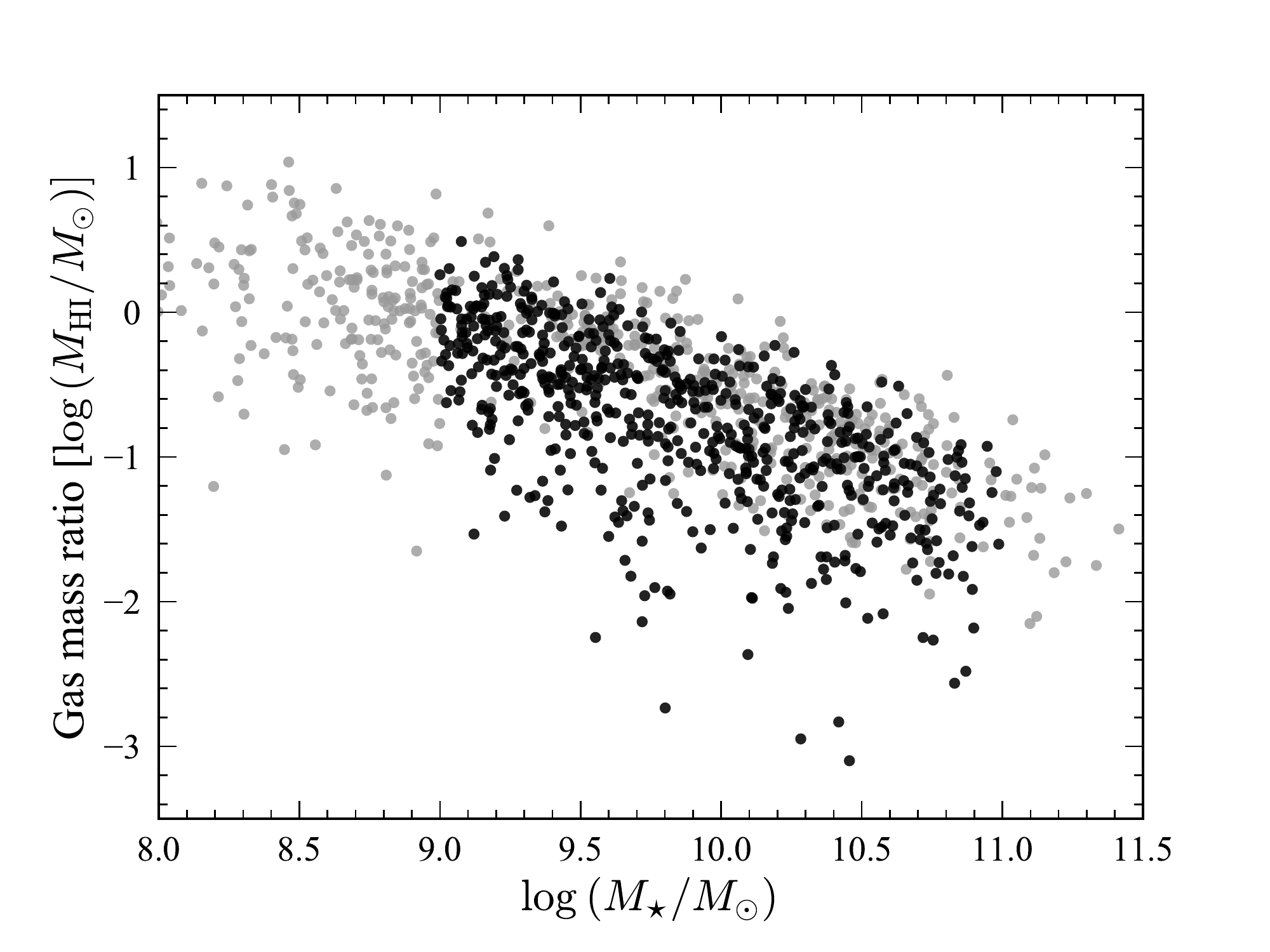}
\end{center}

\caption{Gas mass fraction as a function of stellar mass for all \sfourg{} spiral
galaxies (grey points) and for just the Parent Spiral Sample (black points). 
\label{fig:app-fgas-vs-mstar}}
\end{figure}

Figure~\ref{fig:app-fgas-vs-mstar} shows the correlation between gas mass fraction
and stellar mass for the \sfourg{} spiral galaxies.

\bsp	
\label{lastpage}
\end{document}